\documentclass[aps,prb,twocolumn,showpacs,superscriptaddress,longbibliography,normalem]{revtex4-2}
\usepackage{amsfonts}
\usepackage{amsmath}
\usepackage{amssymb}
\usepackage{graphicx}
\usepackage{float}
\usepackage{multirow}
\usepackage{gensymb}
\usepackage{dcolumn}
\usepackage{ulem}
\usepackage[dvipsnames]{xcolor}

\newcommand{\fgt}{Fe$_3$GeTe$_2$}

\usepackage[bookmarks=false]{hyperref} 
\hypersetup{colorlinks=true,
            linkcolor=blue,
            citecolor=blue,
            urlcolor=orange}
\begin{document}
\title{
Suppression of local magnetic moment formation and paramagnetic exchange interactions in monolayer Fe$_3$GeTe$_2$}
\author{A. A. Katanin}
\affiliation{Center for Photonics and 2D Materials, Moscow Institute of Physics and Technology, Institutsky lane 9, Dolgoprudny, 141700, Moscow Region, Russia}
\affiliation{M. N. Mikheev Institute of Metal Physics of Ural Branch of Russian Academy of Sciences, S. Kovalevskaya Street 18, 620990 Yekaterinburg, Russia}
\author{A. N. Rudenko}
\affiliation{Radboud University, Institute for Molecules and Materials, Heijendaalseweg 135, 6525AJ Nijmegen, The Netherlands}
\author{D. I. Badrtdinov}
\affiliation{Radboud University, Institute for Molecules and Materials, Heijendaalseweg 135, 6525AJ Nijmegen, The Netherlands}
\author{M. I. Katsnelson}
\affiliation{Radboud University, Institute for Molecules and Materials, Heijendaalseweg 135, 6525AJ Nijmegen, The Netherlands}
\begin{abstract}
    We study the electronic and magnetic properties of monolayer Fe$_3$GeTe$_2$ within the DFT+DMFT approach in the paramagnetic phase.
    We argue that this compound is sufficiently far from the local magnetic moment limit, demonstrating non-linear temperature dependencies of the partial inverse local and uniform magnetic susceptibilities in a broad temperature range. We find that in the regime of moderate Coulomb interactions ($U=3-4$~eV), the iron atoms located above and below the Ge plane carry a substantial local magnetic moment ($\mu \gtrsim 4.5 \mu_B$), while the iron atom located within the Ge plane does not exhibit any pronounced magnetic moment. At the same time, the 
    exchange interaction between these two symmetry-nonequivalent types of atoms turn out to be crucial for stabilizing long-range ferromagnetic order in Fe$_3$GeTe$_2$. The estimated spin-wave stiffness and Curie temperature are in good agreement with the experimental data, indicating that a dynamical treatment of electron correlations in Fe$_3$GeTe$_2$ is essential to properly describe its partially itinerant magnetic behavior.
\end{abstract}.
\maketitle

\section{Introduction}
The interest to two-dimensional (2D) materials is driven by the progress in the development of novel technologies involving miniaturization of electronic devices and low-energy consumption. The discovery of 2D magnetic materials \cite{2dmag_1,2dmag_2,2dmag_3,2dmag_4,Gibertini2019,Soriano2020} has further enhanced interest in the field of 2D materials, opening new ways to control magnetism in 2D, which is promising for applications as well as for fundamental research.

Fe$_3$GeTe$_2$ is a typical representative of 2D easy-axis ferromagnets with relatively high Curie temperature, reaching 130~K in the monolayer limit \cite{Deng2018,Fei2018}. 
Unlike many other 2D magnets, \fgt{} is metallic, which gives rise to a variety of conventional transport phenomena \cite{Kim2018,Xu2019,Roemer2020} as well as to more exotic phenomena, such as the Kondo effect \cite{Zhang2018,Mengting2021}. 
On the other hand, the presence of charge carriers challenges the description of magnetic properties in terms of spin models, which hinders understanding of physical mechanisms behind ferromagnetism in \fgt{}. Although the electron-magnon interaction is not expected to be large in \fgt{} \cite{Danis2023}, electronic correlations have been shown to play a considerable role in \fgt{}. This conclusion was based on density functional plus dynamical mean-field theory (DFT+DMFT) \cite{Anisimov1997,Lichtenstein1998,Kotliar2006} calculations performed in the works  \cite{Kim2018,Zhu2016,Kim2022,Ghosh2023,Sharma2024}. Most of these works are focused on quasiparticle spectra probed by photoemission, but Ref. \cite{Ghosh2023} also studies exchange interactions by combining DFT+DMFT with the magnetic force theorem  \cite{Katsnelson2000,Kvashnin2015,Szilva2023}. 

In some of the abovementioned DFT+DMFT studies, the signs of Hund metal behavior in \fgt{} were obtained  \cite{Kim2022,Sharma2024}. In particular, the enhancement of quasiparticle mass and electron damping, Curie behavior of local magnetic susceptibility, and spin-orbital separation were discussed. At the same time, \fgt{} is quite far from the fully localized limit. Indeed, the local magnetic moment $\mu_C\simeq 4.6 \mu_B$/Fe extracted from the Curie-Weiss law and the saturation moment $\mu_s\simeq 1.6\mu_B$/Fe (see, e.g., Refs. \cite{Chen2013,Zhu2016}) yield the ratio of the moment $p_C\simeq 1.9$ extracted from the Curie-Weiss law $ \mu_C^2=4p_C(p_C+1) \mu_B^2$ to the ordering moment $p_s=\mu_s/(2\mu_B)\simeq 0.8$, equal to $p_C/p_s\simeq 2.4$, which for the Curie temperature $T_C\simeq 220$~K corresponds to more or less the same position on the Rhodes-Wohlfarth curve \cite{Rhodes1963,Moriya1985,Santiago2017} as Pd-Cu and Pd-Fe alloys which are substantially far from the localized limit.
Therefore, \fgt{} combines both, the local magnetic moment and itinerant behavior. This is partly due to site-differentiated behavior, discussed earlier in Ref. \cite{Kim2022}, but as we argue in the present paper, we expect much stronger site differentiation, than considered previously.

To describe correlation effects within DMFT 
in the previous study of Ref. \cite{Ghosh2023}, the perturbation solver SPTF (spin-polarized T-matrix plus FLEX) \cite{SPTF1,SPTF2} was used. 
It seems to be reasonable at semi-quantitative level for elemental 3$d$ metals \cite{Braun2006,DiMarco2009,Braun2010} but its applicability for \fgt{} is not obvious. There are also other issues to be clarified. 
The value $U \approx 5$ eV used in the previous calculations \cite{Zhu2016,Kim2018,Kim2022,Ghosh2023,Sharma2024} may be unrealistically large. {This value is based solely on a method applied to an iron compound BaFe$_2$As$_2$ within the self-consistent $GW$ scheme \cite{Kotliar2010}. At the same time, this result appears to be much larger than that obtained within a non-self-consistent $GW$ calculation \cite{Kotliar2010}, as well as from the constrained random phase approximation (cRPA) \cite{Werner2012}. Apart from the weak Coulomb screening, 
some earlier DMFT studies on \fgt{} used a nominal form of double counting (see, e.g., Refs. \cite{Zhu2016,Kim2022}), which fixes the $d$-electron concentration at iron atoms to be close to $n_d=6$. As will be shown below, this approximation is not fully justified for \fgt{}.

In the present paper, we perform a systematic DFT+DMFT study of magnetic interactions and finite-temperature magnetic properties of monolayer \fgt{} focusing on its itinerant magnetic behavior. 
A common way to describe magnetic interactions in correlated systems is to make use of an adapted version of the magnetic force approach \cite{Katsnelson2000,Kvashnin2015,Szilva2023}, which starts from a magnetically ordered state, but {may be limited in a situation with partially formed local magnetic moments \cite{Katsnelson2004,Solovyev2021,Solovyev2024}. Therefore} here we use an alternative formalism based on nonuniform magnetic susceptibility, which is also applicable in the paramagnetic phase \cite{Katanin2023FeNi,Katanin2025SU2}. Earlier, we used this formalism to describe the magnetic properties of iron, nickel \cite{Katanin2023FeNi,Katanin2025SU2}, cobalt \cite{Katanin2023Co}, CrO$_2$ \cite{Katanin2024CrO2}, and CrTe$_2$ \cite{Katanin2025CrTe2,CrTe2Erratum}. For strong magnets (in particular, elemental iron), it produces results that only slightly correct previous DFT estimates in the ordered phase \cite{Katanin2023FeNi}. In contrast to the magnetic force approach, this method does not require the assumption of a certain magnetic state; therefore, it provides an unbiased calculation of exchange interactions. Even more importantly for our case, it does not require the existence of well-defined local magnetic moments and should be applicable for itinerant systems far from the localized limit. This method can also trace the temperature evolution of exchange interactions, including self-energy and vertex corrections.

    Based on our DMFT study in the paramagnetic phase, we provide strong evidence on the itinerant nature of magnetism in Fe$_3$GeTe$_2$. In particular, we find the iron atoms located above and below the Ge plane to be more strongly correlated than the iron atoms located within the Ge plane, in agreement with previous considerations \cite{Kim2022}. The temperature dependencies of both local and partial uniform inverse susceptibilities are non-linear over a wide temperature range, in contrast to the elemental iron \cite{Katanin2023FeNi,Katanin2025SU2}. The obtained local magnetic moment, extracted from the Curie-Weiss law for the uniform magnetic susceptibility, and extrapolated to low temperatures, is found to be in agreement with the experimental data for the moderate Coulomb interactions $U=3- 4$~eV. 
From our study in the paramagnetic phase, we find all dominant exchange interactions to be ferromagnetic, except for the interactions between stronger correlated Fe$_1$-Fe$_1$ and Fe$_2$-Fe$_2$ atoms, i.e., those located above and below the Ge plane. {Although the} magnitudes of the exchange interactions are, to a large extent, consistent with previous results in the ferromagnetic phase \cite{Ghosh2023}, {they refer to the state with only partly formed local magnetic moments}. The estimated spin stiffness and Curie temperature are in good agreement with the experimental data. This agreement highlights the importance of itinerant magnetic behavior in \fgt{}, leading to partial suppression of its ferromagnetism. 

The paper is organized as follows. In Sec.~\ref{Sect:Methods}, we briefly discuss the methods of our study. In Sec.~\ref{Sect:Results}, we present the results for the electronic properties (Sec.~\ref{Sect:Electr}), magnetic susceptibilities (Sec.~\ref{Sect:Magn}), exchange interactions (Sec.~\ref{Sect:Exch}), and the Curie temperature (Sec.~\ref{Sect:Curie}). In Sec. \ref{Sect:Concl}, we conclude our paper. 

  

\section{Methods}
\label{Sect:Methods}

\subsection{DFT details}

\vspace{-.3cm}
Density-functional (DFT) band-structure calculations were performed within the Perdew-Burke-Ernzerhof (PBE) exchange-correlation functional~\cite{PBE} as implemented in  {Vienna ab initio simulation package} ({\sc vasp}) \cite{VASP1,VASP2}. In all DFT calculations, we use a ($18\times18\times1$) $\Gamma$ centered grid for the Brillouin zone integration and set the energy cutoff of the plane-wave basis to 400 eV, the energy convergence criteria to $10^{-8}$ eV. The experimental crystal structure of bulk \fgt{} was used~\cite{Deiseroth2006} with the lattice constant $a = 4$~\AA, in which a vacuum space about 20 \AA\, between the monolayer replicas in the vertical direction was introduced. The positions of atoms were relaxed until all the residual force components on each atom were less than 10$^{-3}$ eV/\AA. From the nonmagnetic electronic structure, maximally localized Wannier functions~\cite{marzari1997}  were constructed using the {\sc wannier90} package~\cite{pizzi2020} projecting onto the $3d$ and $5p$ states of iron and tellurium, respectively. The obtained Wannier functions served as a basis for the tight-binding Hamiltonian, subsequently used in DMFT calculations.

\subsection{DMFT}

In DMFT calculations, we consider the density-density interaction matrix, parameterized by Slater parameters $F^0$, $F^2$, and $F^4$, expressed through Hubbard $U$ and Hund $J_H$ interaction parameters according to 
${F^0\equiv U}$ and ${(F^2+F^4)/14 \equiv J_{\rm H}}$, $F_2/F_4\simeq 0.63$ (see Ref.~\onlinecite{u_and_j}). Since our Hamiltonian does not contain high-energy $s$, $p$ states of iron, we consider moderate $U=2$ to $4$~eV, assuming that it is partly screened by the high-energy states. We also use Hund exchange $J_H=0.9$~eV. We use a double-counting correction ${H}_{\rm DC} = M_{\rm DC}\sum_{ir}  {n}_{ird} $ in the around mean-field form~\cite{AMF}, $M_{\rm DC}=\langle {n}_{ird} \rangle [U (2 n_{\rm orb} {-} 1) - J_{\rm H}  (n_{\rm orb} {-} 1)] / (2 n_{\rm orb})$, where
${n}_{ird}$ is the {operator of the} number of $d$ electrons at the site $(i,r)$, where $i$ is the unit cell index and $r$ is the site index within the unit cell. We perform the Hamiltonian rotation in the $d$-orbital space to diagonalize the crystal field, which considerably reduces the off-diagonal components of the local Green's functions with respect to the orbital indexes and improves applicability of the density-density interaction.

To determine the exchange interactions we consider the effective Heisenberg model with the Hamiltonian $H=-(1/2)\sum_{{\bf q},rr'} J^{rr'}_{\bf q} {\mathbf S}^r_{\mathbf q} {\mathbf S}^{r'}_{-{\mathbf q}}$, {$\mathbf S^r_{\mathbf q}$ is the Fourier transform of static operators ${\mathbf S}_{ir}$},
where the orbital-summed on-site static spin operators ${\mathbf S}_{ir}=\sum_m {\mathbf S}_{irm}$ and ${\mathbf S}_{irm}=(1/2)\sum_{\sigma\sigma'\nu}c^+_{irm\sigma\nu}\mbox {\boldmath $\sigma $}_{\sigma\sigma'}c_{irm\sigma'\nu}$  
is the electron spin operator, $\nu$ are the Matsubara frequencies, $c^+_{irm\sigma\nu}$ and $c_{irm\sigma\nu}$ are the frequency components of the electron creation and destruction operators at the site $(i,r)$, $d$-orbital $m$, and spin projection $\sigma$, and $\mbox {\boldmath $\sigma $}_{\sigma \sigma'
}$ are the Pauli matrices.

To extract the exchange parameters $J_{\bf q}$, we relate them to the orbital-summed non-local static {longitudinal} susceptibility $\chi^{rr'}_{\mathbf q}=-\langle \langle S^{z,r}_{\mathbf q}|S^{z,r'}_{-{\mathbf q}}\rangle\rangle_{\omega=0}=\sum_{mm^{\prime}}\hat{\chi}_{\bf q}^{mr, m^{\prime}r'}$ (the hats stand for matrices with respect to orbital and site indexes; $\langle \langle ..|..\rangle\rangle_\omega$ is the retarded Green's function), considering the generalization of the approach of Ref. \onlinecite{Katanin2023FeNi} to several atoms in the unit cell, and express exchange interactions as 
\begin{equation}
J_{\mathbf q}=
\chi_{\rm loc}^{-1}-\chi_{\bf q}^{-1},
\label{JqAvDef}
\end{equation}
 the matrix inverse in Eq. (\ref{JqAvDef}) is taken with respect to the site indexes in the unit cell. The matrix of local susceptibilities $\chi^{rr'}_{\rm loc}=-\langle \langle S^z_{ir}|S^z_{ir}\rangle\rangle_{\omega=0}\delta_{rr'}=\sum_{{  m}{  m}'} \hat{\chi}^{{  m}{  m}',r}_{\rm loc}\delta_{rr'}$ is diagonal with respect to the site indices. The non-local susceptibility is determined from the Bethe-Salpeter equation using the local particle-hole irreducible vertices.  The latter vertices are extracted from the inverse Bethe-Salpeter equation applied to the particle-hole vertex obtained within the DMFT solution (cf. Refs.~\cite{Katanin2023FeNi,OurRev}).

The DMFT calculations of the self-energies, non-uniform susceptibilities, and exchange interactions were performed within the Wan2mb software package \cite{Katanin2023FeNi,Katanin2023Co,Katanin2024CrO2,Katanin2025CrTe2,katanin2026hubbdmft}, based on the continuous-time hybridization expansion Quantum Monte Carlo (CT-QMC) method for solving the impurity problem \cite{Werner2006}, implemented in the iQIST software package \cite{iQIST}.

\section{Results}
\label{Sect:Results}

\subsection{Electronic properties}
\label{Sect:Electr}

\begin{table}[b]
\begin{tabular}{||l|c|c|c|c|c||}
\hline\hline
& Atomic & L\"owdin & \multicolumn{3}{c||}{Wannier} \\ \hline
& \multicolumn{1}{l}{} & \multicolumn{1}{l|}{} & $U=0$ & $U=2$& $U=4$\\ \hline
Fe$_{1,2}$ & \multicolumn{1}{c|}{6.22} & \multicolumn{1}{c|}{6.92} & 
\multicolumn{1}{l|}{7.48} & \multicolumn{1}{l|}{6.87} & 
\multicolumn{1}{l||}{6.54} \\ \hline
Fe$_{3}$ & \multicolumn{1}{c|}{6.45} & \multicolumn{1}{c|}{6.98} & 
\multicolumn{1}{l|}{7.65} & \multicolumn{1}{l|}{7.91} & 
\multicolumn{1}{l||}{7.81} \\ \hline\hline
\end{tabular}
\caption{Occupations of $d$-states at $\beta=10$~eV$^{-1}$ in various approaches (see text). The values of $U$ are in units of eV.}
\label{Tableocc}
\end{table}

\begin{figure}[t]
		\center{		\includegraphics[width=1.0\linewidth]{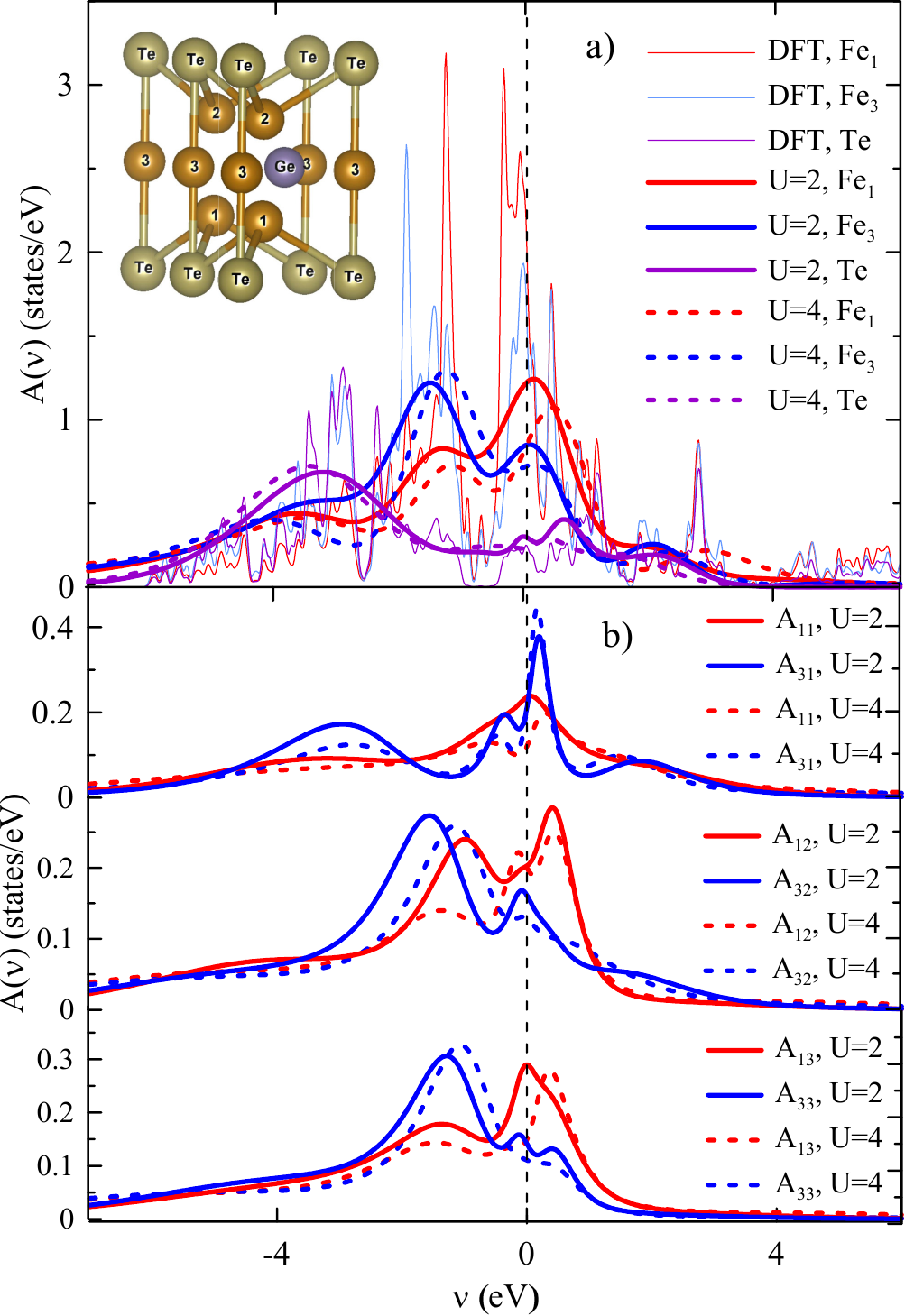}}
		\caption{a) Frequency dependence of partial densities of $d$ states of iron  and p states of tellurium calculated within the   DFT+DMFT approach at $\beta=10$~eV$^{-1}$ and $U=2$~eV and $4$~eV. The inset shows the crystal structure of \fgt{}, numbers $r=1..3$ correspond to Fe$_r$ atoms in the unit cell. b) Orbital-resolved contributions $A_{rm}(\nu)$ to the density of states of $r$-th atom from $m$-th Fe orbital in DFT+DMFT approach. Vertical dashed line shows the position of the Fermi level.}
\label{Fig_DOS}
\end{figure}

\begin{figure}[t]
		\center{		\includegraphics[width=0.95\linewidth]{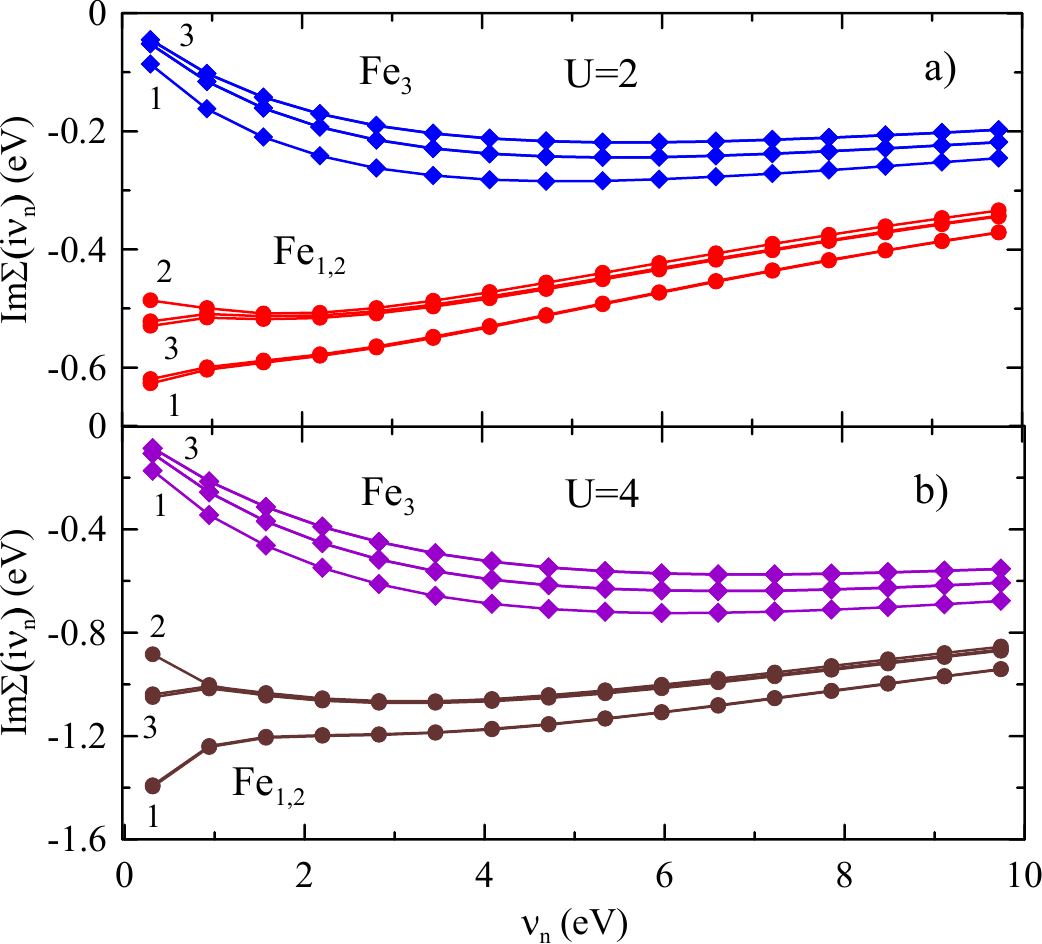}}
		\caption{Frequency dependence of the imaginary parts of the electronic self-energies, corresponding to various $d$-orbital states at Fe$_{1,2}$ atoms (circles) and Fe$_3$ atoms (diamonds) calculated within the   DFT+DMFT approach at $\beta=10$~eV$^{-1}$ and $U=2$~eV (a), $U=4$~eV (b). Numbers at the curves refer to the orbital state (see text). }
\label{Fig_Sigma}
\end{figure}

The occupation of $d$ states of non-equivalent Fe atoms is quite different (see Table \ref{Tableocc}, throughout the paper, we denote Fe$_{1,2}$ as the atoms located above (below) the plane, and Fe$_3$ denote the atoms located in the plane). While the occupations calculated from the integration over the atomic spheres are close to $6$ (see also Ref. \cite{Kim2022}), the L\"owdin charges and Wannier states occupations (which also account for the charge outside the atomic spheres) are higher, showing potential insufficiency of the nominal occupations analysis \cite{Kim2022}. Although the occupations of Fe$_{1,2}$ $d$ states (extracted from Wannier orbitals) are somewhat decreased by the interaction (yielding their occupation somewhat closer to half filling), the occupation of the $d$-shell of Fe$_3$ atom even increases.  

In Fig. \ref{Fig_DOS}(a), we present the partial densities of states of iron and tellurium atoms in DFT and DFT+DMFT approaches. The peak of the density of states at the Fermi level is formed by the $d_{xy}$ and $d_{x^2-y^2}$ orbital contributions of both, Fe$_{1,2}$ and Fe$_3$ atoms. One can see that correlations yield a shift of the peak of the density of states of iron atoms to larger energies (in particular, pushing the peak from the Fermi level to a position above the Fermi level), while the profile of the density of states of tellurium atoms shifts to lower energies. This leads to the above mentioned decrease of the occupation numbers of Fe$_{1,2}$ atoms by correlations,
while the occupation numbers of tellurium atoms increase.  


\begin{table}[b]
\begin{tabular}{||l|l|l|l|l|l||}
\hline\hline
& 1 & 2 & 3 & 4 & 5 \\ \hline
Fe$_{1,2}$ & $d_{x^{2}-y^{2}};d_{yz}$ & $d_{3z^{2}-r^{2}}$ & $%
d_{x^{2}-y^{2}};d_{yz}$ & $d_{xz};d_{xy}$ & $d_{xz};d_{xy}$ \\ \hline
Fe$_{3}$ & $d_{3z^{2}-r^{2}}$ & $d_{x^{2}-y^{2}}$ & $d_{yz}$ & $d_{xz}$ & $%
d_{xy}$ \\ \hline\hline
\end{tabular}
\caption{Orbital content of the states, obtained by diagonalization of the crystal field}
\label{TableOrb}
\end{table}

In Fig. \ref{Fig_DOS}(b) we present the orbital-resolved contributions to iron densities of states. The orbital content of various states, obtained by the diagonalization of crystal field, is listed in Table \ref{TableOrb}. 
The partial densities of states 4 and 5
are not shown since they coincide with those for states 1 and 3 for Fe$_{1,2}$ atoms and 
states 3 and 2 for Fe$_{3}$ atom. One can see that the correlations suppress the density of states of all the orbitals  of Fe$_{1,2}$ except orbital 2, which is related to the enhancement of correlation effects by the peaks of the bare density of states of $d_{xy}$ and $d_{x^2-y^2}$ orbitals (contributing to orbitals 1,3-5, as discussed above). 

In Fig. \ref{Fig_Sigma}, we present the electronic self-energies. One can see that, in accordance with the above discussed suppression of the partial densities of states of Fe$_{1,2}$ atoms, the respective self-energies of the orbitals 1,3-5 have a non-quasiparticle form with $\partial {\rm Im}\Sigma(i\nu_n)/\partial \nu_n>0$ (which is related to the formation of the local magnetic moment, discussed below), while the self-energies of the atoms Fe$_{3}$ have a quasiparticle form. The violation of quasiparticle form of the orbital states $1,3,4,5$ of Fe$_{1,2}$ atoms is also related to their enhanced bare density of states, similarly to the $e_g$ studies of elemental iron \cite{Katanin2010AlphaIron}. For the states 1,4 it is also accompanied by closer proximity to half filling ($n_{1,1(4)}\simeq 0.6$ in comparison to the occupations of the other states ($n_{1m}\simeq 0.7$, $m\ne 1,4$) of the same atoms.

\subsection{Local and uniform susceptibilities}
\label{Sect:Magn}

\begin{figure}[b]
		\center{		\includegraphics[width=1.0\linewidth]{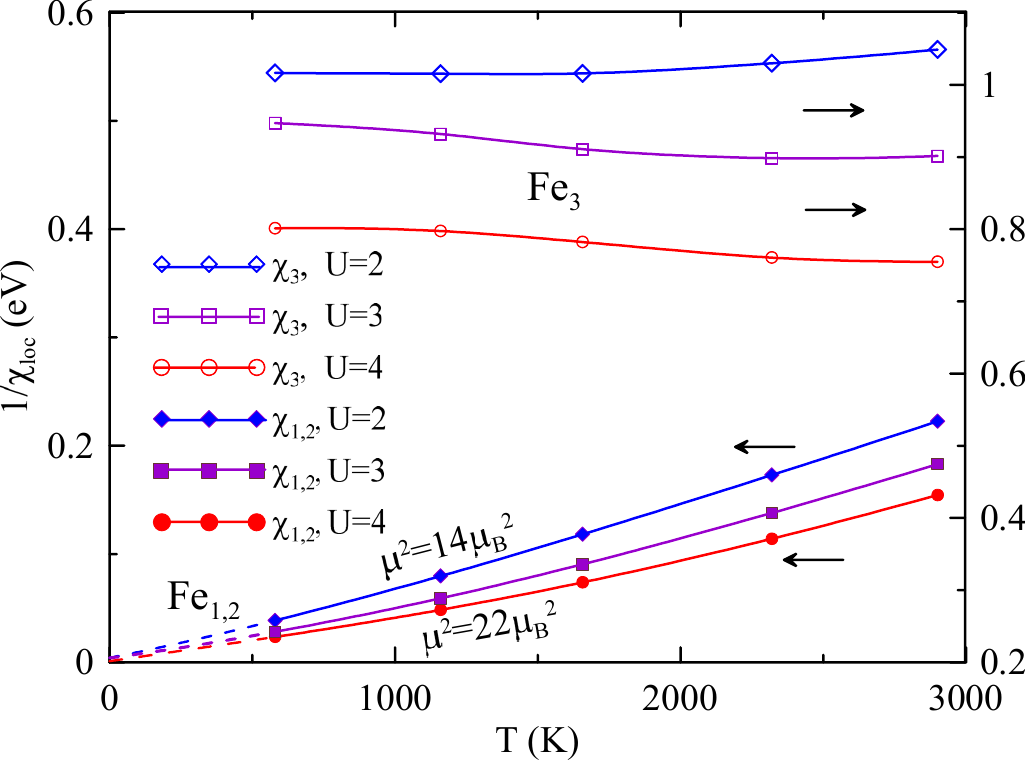}}
		\caption{
Temperature dependence of the inverse local  susceptibilities of Fe$_{1,2}$ (left axis) and Fe$_3$ (right axis) atoms calculated within the DFT+DMFT approach. Dashed lines show the result of extrapolation. The values of $U$ are in units of eV.}
\label{Fig_chiloc}
\end{figure}

Fig. \ref{Fig_chiloc} shows the temperature dependence of the local susceptibilities. In accordance with the self-energies obtained, the local susceptibilities at the Fe$_{1,2}$ atoms show Curie-like behavior, corresponding to the formation of local magnetic moments, while Fe$_{3}$ atoms exhibit Pauli-like susceptibilities. However, we note that the inverse susceptibilities corresponding to the Fe$_{1,2}$ atoms are not perfectly linear, indicating that the formation of local moments is only partial. Moreover, the size of the local magnetic moment, determined by the slope of $\chi_{\rm loc}^{-1}(T)$ dependencies, increases with the interaction due to closer proximity to half filling of Fe$_{1,2}$ atoms at larger $U$. 
At $U=3-4$~eV and the relatively high temperature considered, we obtain a magnetic moment $\mu^2_{\rm loc}\simeq 20\mu_B^2$/Fe$_{1,2}$. Assuming that the third atom does not carry a pronounced local magnetic moment, we therefore obtain $\mu^2_{\rm loc}\simeq 40\mu_B^2$, which is smaller than  $\mu^2\simeq 63\mu_B^2$ extracted experimentally from the Curie-Weiss law for the {\it uniform} susceptibility \cite{Kim2022,Sharma2024}.
{The difference between the calculated $\mu^2_{\rm loc}$ and the experimental value} can be partially attributed to the relatively high temperatures considered. By performing an extrapolation with a polynomial fit to $T\sim T_C^{\rm exp}\simeq 200$~K, we find a larger magnetic moment of $\mu_{\rm loc}^2\simeq 60\mu_B^2$/Fe$_{1,2}$ in better agreement with the experimental data. 

\begin{figure}[t]
		\center{		\includegraphics[width=1.0\linewidth]{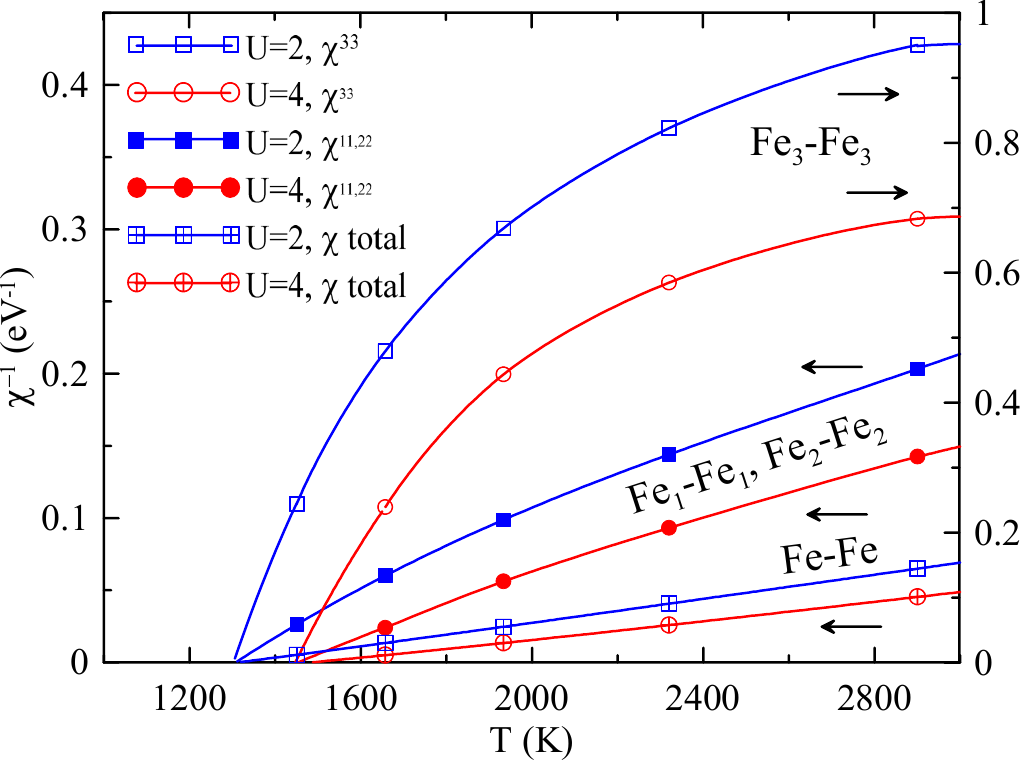}}
		\caption{Temperature dependence of the inverse partial (open and closed symbols) and total (crossed symbols) uniform spin susceptibilities of Fe atoms calcualted within the DFT+DMFT approach with $U=2$~eV (squares) and $U=4$~eV (circles). Solid lines show the result of interpolation.}
\label{Fig_chiT}
\end{figure}

The abovementioned theoretical values of magnetic moments, obtained from local magnetic susceptibility, are also close to those extracted from
the uniform susceptibility measured experimentally.
Fig. \ref{Fig_chiT} shows the temperature dependencies of the inverse uniform partial $(\chi^{mm}_{{\mathbf q}=0})^{-1}$ (corresponding to Fe$_m$ atom) and total $(\sum_{mn}\chi^{mn}_{{\mathbf q}=0})^{-1}$  susceptibilities. The inverse susceptibilities corresponding to the atoms Fe$_{1,2}$ show a Curie-Weiss-like behavior, with moderate non-linearity corresponding to the partial formation of local magnetic moments, while the inverse susceptibilities corresponding to Fe$_{3}$ atoms show linear behavior only at sufficiently high temperatures, which changes to non-linear behavior at lower temperatures. Interestingly, the total inverse susceptibilities show a more linear behavior, which is due to the important contribution of the off-diagonal susceptibility elements $\chi^{mn}_{{\mathbf q}=0}$ with $m\ne n$. The Curie temperature obtained within DMFT is $T_C^{\rm DMFT}\simeq 1300$~K for $U=2$~eV and $T_C^{\rm DMFT}\simeq 1450$~K for $U=4$~eV. The total square of the magnetic moment estimated in the vicinity of $T_C^{\rm DMFT}$ is $\mu^2\simeq 13\mu_B^2$/Fe ($9\mu_B^2$/Fe) for $U=4$~eV ($2$~eV). 

We note that due to the mean-field nature of the approach, DMFT violates the Mermin-Wagner theorem and yields a finite Curie temperature $T_C^{\rm DMFT}$, which physically corresponds to the crossover to the regime of strong magnetic correlations. Therefore, the actual Curie temperature is much smaller than those obtained from the DMFT and is estimated below in Sect. \ref{Sect:Curie}. Extrapolating the obtained temperature dependencies of magnetic moments to the low-temperature region, we find at $T\sim T_C^{\rm exp}$ the estimate $\mu^2\simeq 21 \mu_B^2$/Fe ($17 \mu_B^2$/Fe) for $U=4$~eV ($2$~eV), respectively. The latter value agrees with that obtained experimentally. Therefore, despite the presence of weakly magnetic Fe$_3$ atoms, which do not possess well formed local magnetic moments, the considered magnetic state of the atoms is sufficient to explain the experimentally measured local magnetic moment.

\subsection{Exchange interactions and spin-wave dispersions}
\label{Sect:Exch}

\begin{figure}[b]
		\center{		\includegraphics[width=1.0\linewidth]{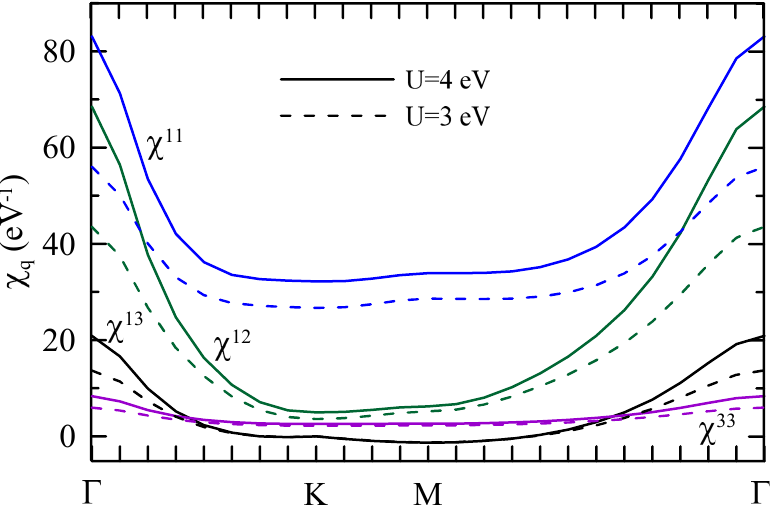}}
		\caption{Momentum dependence of inequivalent nonuniform spin susceptibilities $\chi^{rr'}_{\mathbf q}$ of Fe atoms calcualted within the DFT+DMFT approach with $\beta=7$~eV$^{-1}$, $U=4$~eV (solid lines) and $U=3$~eV (dashed lines).}
\label{Fig_chiq}
\end{figure}

\begin{figure}[t]
		\center{	\includegraphics[width=0.95\linewidth]{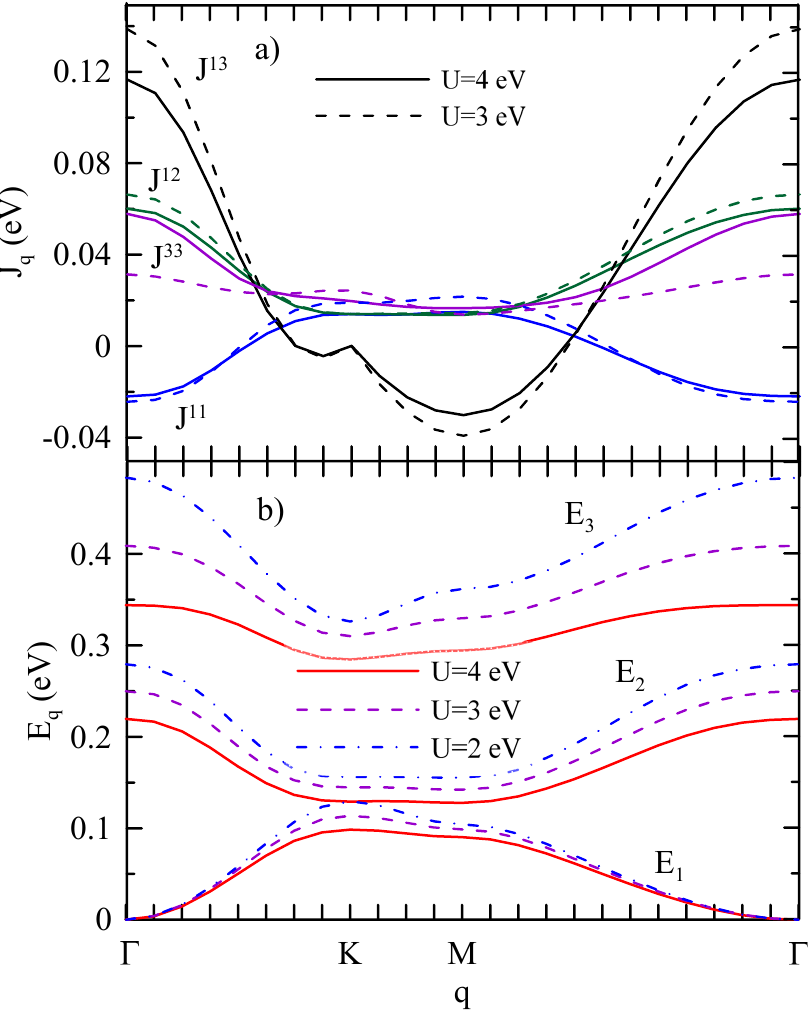}}
		\caption{
(a) Momentum dependence of inequivalent exchange interactions $J_{\mathbf q}^{rr'}$ between various iron atoms $r,r'=1..3$ and (b) magnon dispersions $E_i(q)$ at $\beta=10$~eV$^{-1}$ and various values of the on-site Coulomb interaction $U$.}
\label{Fig_Jq}
\end{figure}

Our method of calculating the exchange interactions from the inverse susceptibilities allows us to consider the situation with no well defined local magnetic moment on Fe$_3$ atoms. 
In Figs.~\ref{Fig_chiq} and~\ref{Fig_Jq}, we show the obtained momentum dependence of susceptibilities and exchange interactions for $U=3$~eV and $U=4$~eV. The components of the susceptibilities have maxima at the $\Gamma$ point, showing the tendency toward ferromagnetic order. At the same time, the exchange interactions $J^{11,22}$ (which are determined by the inverse susceptibilities matrix) have a minimum at the $\Gamma$ point, which corresponds to {\it antiferromagnetic} exchange between the respective atoms, in agreement with previous DFT and DFT+DMFT analysis in the ordered phase \cite{Ghosh2023}. Importantly, these exchange interactions are compensated by other types of interactions, which are all ferromagnetic (with the largest $J^{r3}$ interaction, $r=1,2$). Although this compensation was considered previously within the ordered state (see, e.g., Ref. \cite{Ghosh2023}), here it is obtained for the paramagnetic state, in which one of the three atoms (Fe$_3$) {does not} possess a well defined local magnetic moment. As we argue in Appendix, $J^{r3}$ interactions between more localized Fe$_{1,2}$ and itinerant Fe$_3$ sites
induce the RKKY exchange interactions between local magnetic moments at Fe$_{1,2}$ 
sites. While the exchange interaction $J_{\mathbf q}^{33}$ between Fe$_3$ atoms increases with increasing $U$, the off-diagonal components $J_{\mathbf q}^{12}$ and $J_{\mathbf q}^{13}$ decrease with increasing $U$, which may be related to the redistribution of electronic density towards the Te atoms discussed in Sec.~\ref{Sect:Electr}, and therefore, weaker hybridization between the iron atoms. 

\begin{figure}[t]
		\center{\includegraphics[width=0.9\linewidth]{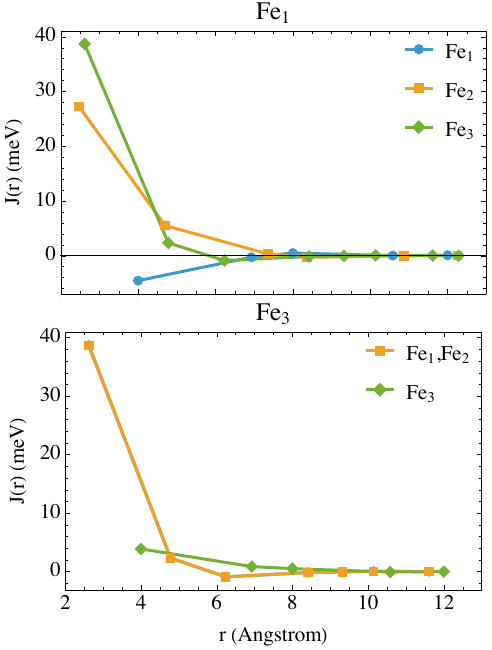}}
		\caption{
Exchange interactions $J^{rr'}({r})$ of Fe$_1$ ($r=1$, top) and Fe$_3$ ($r=3$, bottom) atoms with various neighbors at $U=4$~eV, $\beta=10$~eV$^{-1}$.}
\label{Fig_J}
\end{figure}

The radial $J^{rr'}({r})$ distribution of exchange interactions, obtained by performing the Fourier transform of Eq.~(\ref{JqAvDef}), is shown for $U=4$~eV in Fig. \ref{Fig_J}. The exchange interactions are to a large extent similar to those determined in the ferromagnetic phase within the DFT+DMFT approach \cite{Ghosh2023}. In particular, we find $J^{11,22}<0$ at the nearest neighbor sites in line with the above consideration. The $J^{33}$ interaction at the nearest neighbor sites appears to be ferromagnetic in the present approach, in contrast to the study in the ferromagnetic phase. Also, the $J^{12}$ interaction at the next nearest neighbor sites appears to be ferromagnetic, while it was found to be negligibly small in the ferromagnetic phase. 
Therefore, the present study shows a strong tendency towards ferromagnetism, with the only antiferromagnetic interaction $J^{11}$ at the nearest neighbor sites, the magnitude of which is close to that obtained in Ref. \cite{Ghosh2023}.

To obtain the spin-wave dispersions in the ferromagnetic phase (assuming that the exchange interactions do not change strongly), we use the linear spin-wave approach to the effective Heisenberg model, which yields the spin-wave dispersions as eigenvalues of the Hamiltonian matrix (see, e.g., Ref. \cite{Tyablikov1967})
\begin{equation}
    \mathcal{H}^{rr'}_\mathbf{q} = \delta_{rr'}  \sum_{r''=1}^{3} p_{r''} {J}^{rr''}_0  - \sqrt{p_r p_{r'}} {J}^{rr'}_\mathbf{q}
\end{equation}
where $p_r$ is the ordering moment of site $r$ (in units of $2\mu_B$).
Following the experimental study of Ref. \cite{May2016}, 
we assume 
$p_{1,2}=1.1$, $p_3=0.75$.

\begin{figure}[t]
		\center{\includegraphics[width=0.95\linewidth]{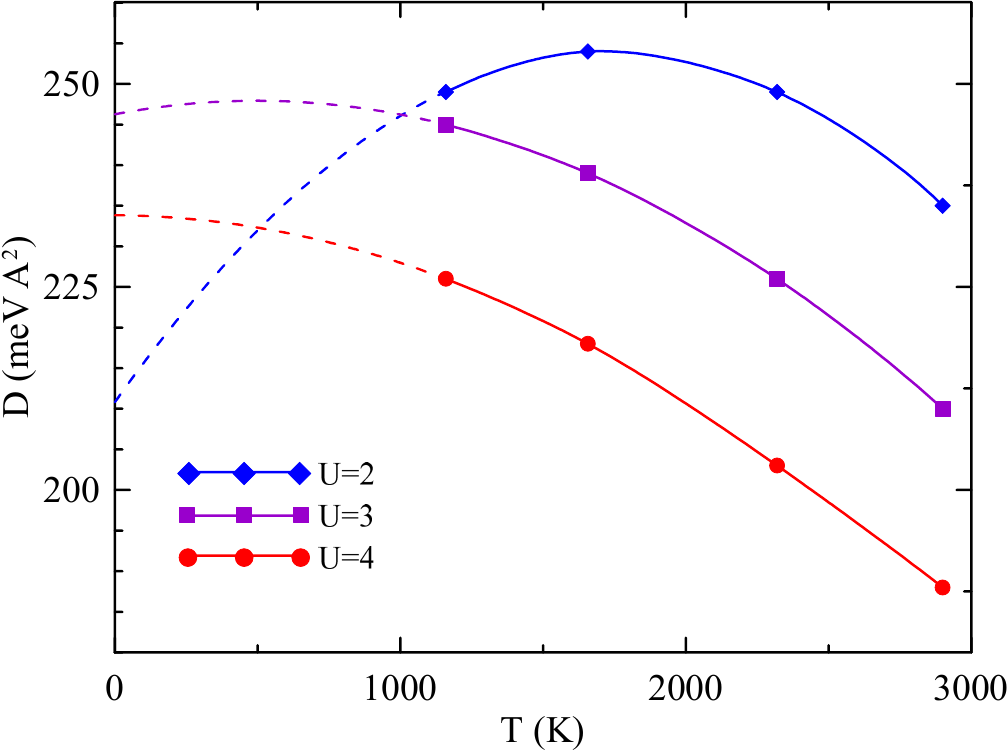}}
		\caption{
Temperature dependence of the spin wave stiffness within the   DFT+DMFT approach for various $U$. Dashed lines show the result of extrapolation. The values of $U$ are in units of eV.}
\label{Fig_D}
\end{figure}

The resulting spin wave dispersions and stiffnesses are shown in Figs. \ref{Fig_Jq}(b) and \ref{Fig_D}, respectively. One can see that the acoustic branch is positively defined. Because of the 
decrease of the off-diagonal components of exchange interactions with increasing $U$, 
both the spin-wave dispersions and the spin stiffnesses decrease with increasing interaction. The resulting spin-wave stiffness $D\simeq$ $235$~meV$\cdot$\AA$^2$ at $U=4$~eV is in {reasonable} agreement with the experimental data $D\simeq 200$~meV$\cdot$\AA$^2$, obtained from the inelastic neutron scattering \cite{Trainer2022}, and is much smaller than the DFT estimate $D\simeq 400$~meV$\cdot$\AA$^2$ \cite{Pushkarev2023}. Note that the overestimate of the spin-wave stiffness may be particularly related to the overestimate of the magnetic moments in Ref. \cite{May2016}, which yield the average saturation moment $\mu_s\simeq 2\mu_B/$Fe larger than the experimental value $\mu_s\simeq 1.6 \mu_B$/Fe \cite{Chen2013}. By rescaling the moments to the values $p_{1,2}=0.9$, $p_3=0.6$ to obtain the total saturation moment $\mu_s=1.6\mu_B$/Fe, we achieve even better agreement with the experimental data, $D=190$~meV~$\cdot$~\AA$^2$. 

\subsection{Curie temperature}
\label{Sect:Curie}

According to the Mermin-Wagner theorem, in an isotropic two-dimensional system, there cannot be long-range magnetic ordering at finite temperatures, that is, 
the Curie temperature $T_C = 0$.
Thus, the finite Curie temperature appears solely due to the  magnetic anisotropy, which can be approximated by the easy-axis term in the spin Hamiltonian $E\sum_{i,r} (S^z_{ir})^2$. The corresponding Curie temperature of a two-dimensional magnet can be estimated from the exchange interactions using the equation \cite{PhysRevB.57.379,PhysRevB.60.1082}
\begin{equation}
     T_{C} = \frac{4 \pi \rho_s}{\ln{({T_{C}}}/{ \Delta})+4\ln ({4\pi \rho_s/T_{C})}}
     \label{EqTC}
\end{equation}
where $\Delta = (2S - 1)E$ is the gap in the spin-wave spectrum, 
$\rho_s = DS/A$ is the spin stiffness, $A$ is the cell area. 
For $E=0.35$~meV (Ref. \onlinecite{Danis2023}), 
$S=1$, $A=13.85$~\AA$^2$, and {$D=235$}~meV$\cdot$ \AA$^2$ ($190$~meV$\cdot$ \AA$^2$), we find $T_C=${170~K} ({140~K}), in a {reasonable} (good) agreement with the experimental data available for monolayer samples (130~K) \cite{Fei2018}, and well below the experimental values for bulk \fgt{} (220~K) \cite{Chen2013}. 


\section{Conclusions}
\label{Sect:Concl}

In conclusion, we have studied the electronic and magnetic properties of \fgt{} within the DFT+DMFT approach in the paramagnetic phase. The obtained temperature dependencies of local and uniform susceptibilities are non-linear, showing only partial formation of local magnetic moments. This can be related to the fact that the $d$-states in \fgt{} are farther away from half filling than in elemental iron. Yet, the iron atoms located above (below) the Ge plane exhibit a non-quasiparticle form of the electronic self-energy.

The partial inverse local and uniform susceptibilities show non-linear temperature behavior, suggesting partially formed magnetic moments. 
The local magnetic moments extracted from the temperature dependence of the local and uniform susceptibilities and extrapolated to the low-temperature region are in reasonable agreement with the experimental data. 
The exchange interactions overall agree with the values obtained earlier within the DFT+DMFT approach in the ferromagnetic phase, but show a lesser degree of frustration.
The spin-wave stiffness ($D\simeq 235$ meV$\cdot$\AA$^2$) and the respective Curie temperature ($T_C=170$~K) are found to be in reasonable agreement with the experimental data.
The overestimate of spin stiffness and Curie temperature is possibly related to the overestimate of saturated magnetic moments; for the total magnetic moment $\mu_s=1.6 \mu_B$/Fe we obtain $D=190$~meV$\cdot$\AA$^2$ and $T_C=140$~K in even better agreement  with the experimental data.

Our results highlight the important relationship between the itinerant behavior of \fgt{} and the suppression of the local magnetic moment at iron atoms located within the Ge plane (Fe$_3$), leading to a strong differentiation of magnetic sites, much larger than previously proposed \cite{Kim2022}. 
Therefore, the approach used here to treat electron correlations allows us to gain further insight into the nature of magnetism of \fgt{}, while its validity is justified by a good agreement of the obtained results with the experimental data.

\vspace{.1cm}

\section*{Acknowledgements}
The work of A.K. was supported by the Russian Science Foundation project 24-12-00186 (performing
DMFT calculations) and the Ministry of Science and Higher education project FSMG-2025-0005 (the development of the DMFT software for multiple non-equivalent atoms in the unit cell).  The authors declare that this work has been published as a result of peer-to-peer scientific collaboration between researchers. The provided affiliations represent the actual addresses of the
authors, in agreement with their digital identifier (ORCID), and cannot be considered as a formal collaboration between Radboud University and the other aforementioned institutions.

\section*{Appendix}
To emphasize the physical meaning of exchange interactions with Fe$_3$ atoms, we consider effective exchange interactions between Fe$_{1,2}$ atoms, for which a partly formed local magnetic moment is obtained by performing $2\times 2$ matrix inversion in Eq. (\ref{JqAvDef}), corresponding to the atoms Fe$_{1,2}$. Using the identity 
$(\chi_{2,2})^{-1}=(\chi^{-1})_{2,2}-(\chi^{-1})_{2,3} (\chi^{-1})_{3,2}/(\chi^{-1})_{3,3}$ where the index 2 corresponds to the Fe$_{1,2}$ atoms, represented by the elements 1,2 of the matrix, and 3 corresponds to Fe$_3$ atom, represented by the third element, we obtain 
\begin{equation}
\tilde{J_{\mathbf q}}=(J_{\mathbf q})_{2,2} + (J_\textbf{q})_{2,3} (\chi_{\mathbf q})_{3,3} (J_\textbf{q})_{3,2}.
\end{equation}
Therefore, the resulting exchange interaction $\tilde{J}_{\mathbf q}$ 
includes the RKKY exchange 
mediated by 
the third atoms, 
coupled by the non-local interactions $J^{r3}$ with $r=1,2$ to local magnetic moments.
We have verified that the spin-wave stiffness, obtained with exchange interactions $\tilde{J}_{\mathbf q}$, is close to that obtained with the full set of exchange interactions, provided that the magnetic moments are renormalized as $\tilde{p}_{1,2}=p_{1,2}+p_3/2$ to include the magnetic moment of the third atom.




\bibliography{main} 


\end{document}